\documentclass[10pt,journal,epsfig]{IEEEtran}
\usepackage{graphicx}
\usepackage{subfigure,color}
\usepackage{amssymb}
\usepackage{cite,comment}
\usepackage{amsmath}
\usepackage{bm,comment}
\usepackage{algorithm}
\usepackage{algpseudocode}
\makeatletter

\newcommand{\Rmnum}[1]{\expandafter\@slowromancap\romannumeral #1@}
\newcommand{\tabincell}[2]{\begin{tabular}{@{}#1@{}}#2\end{tabular}}
\makeatother

\newtheorem{proposition}{\underline{Proposition}}
\newtheorem{remark}{\underline{Remark}}

\newtheorem{lemma}{\underline{Lemma}}

\newtheorem{example}{\underline{Example}}

\begin{document}
\title{Uplink Cooperative NOMA for Cellular-Connected UAV}
\author{Weidong Mei and Rui Zhang, \IEEEmembership{Fellow, IEEE}\vspace{-15pt}
\thanks{\footnotesize{W. Mei is with the NUS Graduate School for Integrative Sciences and Engineering, National University of Singapore, Singapore 119077, and also with the Department of Electrical and Computer Engineering, National University of Singapore, Singapore 117583 (e-mail: wmei@u.nus.edu).}}
\thanks{\footnotesize{R. Zhang is with the Department of Electrical and Computer Engineering, National University of Singapore, Singapore 117583 (e-mail: elezhang@nus.edu.sg).}}}
\maketitle

\begin{abstract}
Aerial-ground interference mitigation is a challenging issue in the emerging cellular-connected unmanned aerial vehicle (UAV) communications. Due to the strong line-of-sight (LoS) air-to-ground (A2G) channels, the UAV may impose/suffer more severe uplink/downlink interference to/from the cellular base stations (BSs) as compared to the ground users. To tackle this challenge, we propose in this paper to apply the non-orthogonal multiple access (NOMA) technique to the uplink communication from a UAV to cellular BSs, under spectrum sharing with the existing ground users. However, for our considered system, traditional NOMA with only local interference cancellation (IC) at individual BSs, termed non-cooperative NOMA, may provide very limited gain compared to the orthogonal multiple access (OMA). This is because there are a large number of co-channel BSs due to the LoS A2G channels and thus the rate performance of the UAV is severely limited by the BS with the worst channel condition with the UAV. To mitigate the UAV's uplink interference without significantly compromising its achievable rate, a new \emph{cooperative NOMA} scheme is proposed in this paper by exploiting the existing backhaul links among BSs. Specifically, some BSs with better channel conditions are selected to decode the UAV's signals first, and then forward the decoded signals to their backhaul-connected BSs for IC. To investigate the optimal design of cooperative NOMA and air-ground performance trade-off, we maximize the weighted sum-rate of the UAV and ground users by jointly optimizing the UAV's rate and power allocations over multiple resource blocks as well as their associated BSs. However, this problem is difficult to be solved optimally. To obtain useful insights, we first consider two special cases with egoistic and altruistic transmission strategies of the UAV, respectively, and solve their corresponding problems optimally. Next, we consider the general case and propose an efficient suboptimal solution by applying the alternating optimization and successive convex approximation techniques. Numerical results show that the proposed cooperative NOMA scheme yields significant throughput gains than the traditional OMA as well as the non-cooperative NOMA benchmark.
\end{abstract}
\begin{IEEEkeywords}
Cell association, cellular-connected UAV, cooperative interference cancellation, non-orthogonal multiple access (NOMA), power control, unmanned aerial vehicle (UAV).
\end{IEEEkeywords}

\section{Introduction}
The demand for unmanned aerial vehicles (UAVs) or drones, has been soaring globally over the recent years, due to their cost effectiveness and capability to perform complex tasks in various applications such as aerial imaging, cargo transport, traffic monitoring, and communication platform\cite{zeng2016wireless}. To pave the way towards the upcoming era of ``internet-of-drones''\cite{gharibi2016internet}, it is imperative to support high-performance UAV-ground communications with ubiquitous coverage, low latency, and high reliability/throughput, in order to realize real-time command and control for UAV safe operation as well as rate-demanding payload data communication with ground users\cite{kopardekar2014unmanned}. However, at present, almost all UAVs in the market communicate with the ground via the simple direct point-to-point links over the unlicensed spectrum (e.g., the industrial, scientific and medical (ISM) bands), which is typically of limited data rate, unreliable, insecure, vulnerable to interference, and can only operate within the visual line-of-sight (VLoS) range, thus severely limiting the future applications of UAVs. Recently, \emph{cellular-connected UAV} has been considered as a promising new solution, by integrating UAVs into the cellular network as new aerial user equipments (UEs) served by the ground base stations (BSs). Thanks to the superior performance of today's Long Term Evolution (LTE) and future fifth-generation (5G) cellular networks, cellular-connected UAV is anticipated to achieve significant performance enhancement over the existing point-to-point UAV-ground communications, in terms of all of reliability, coverage and throughput\cite{muruganathan2018overview}. Preliminary field trials have also demonstrated that it is feasible to support the basic communication requirements for UAVs with the current LTE network\cite{van2016lte,qualcom2017lte}.

Despite the above advantages of cellular-connected UAVs, how to mitigate the severe aerial-ground interference is still considered as a major challenge in enabling the efficient coexistence between existing ground and new aerial UEs. Different from the conventional terrestrial systems, the high UAV altitude leads to unique air-to-ground (A2G) line-of-sight (LoS) dominated channels in cellular-connected UAV communication, which bring both opportunities and challenges. On one hand, the presence of LoS links leads to more reliable communication channels as compared to terrestrial channels between the ground UEs and BSs, which in general suffer from more severe path-loss, shadowing and multi-path fading. Besides, the LoS channels also make a UAV being potentially served by much more BSs at the same time, thus yielding a higher macro-diversity gain in BS associations compared to ground UEs. However, on the other hand, the dominance of LoS links also renders the UAV to impose/suffer more severe uplink/downlink interference to/from a much larger number of BSs than ground UEs, which may significantly degrade the communication performance of UAVs in the downlink as well as ground UEs in the uplink.

\begin{figure}[!t]
\centering
\includegraphics[width=3.4in]{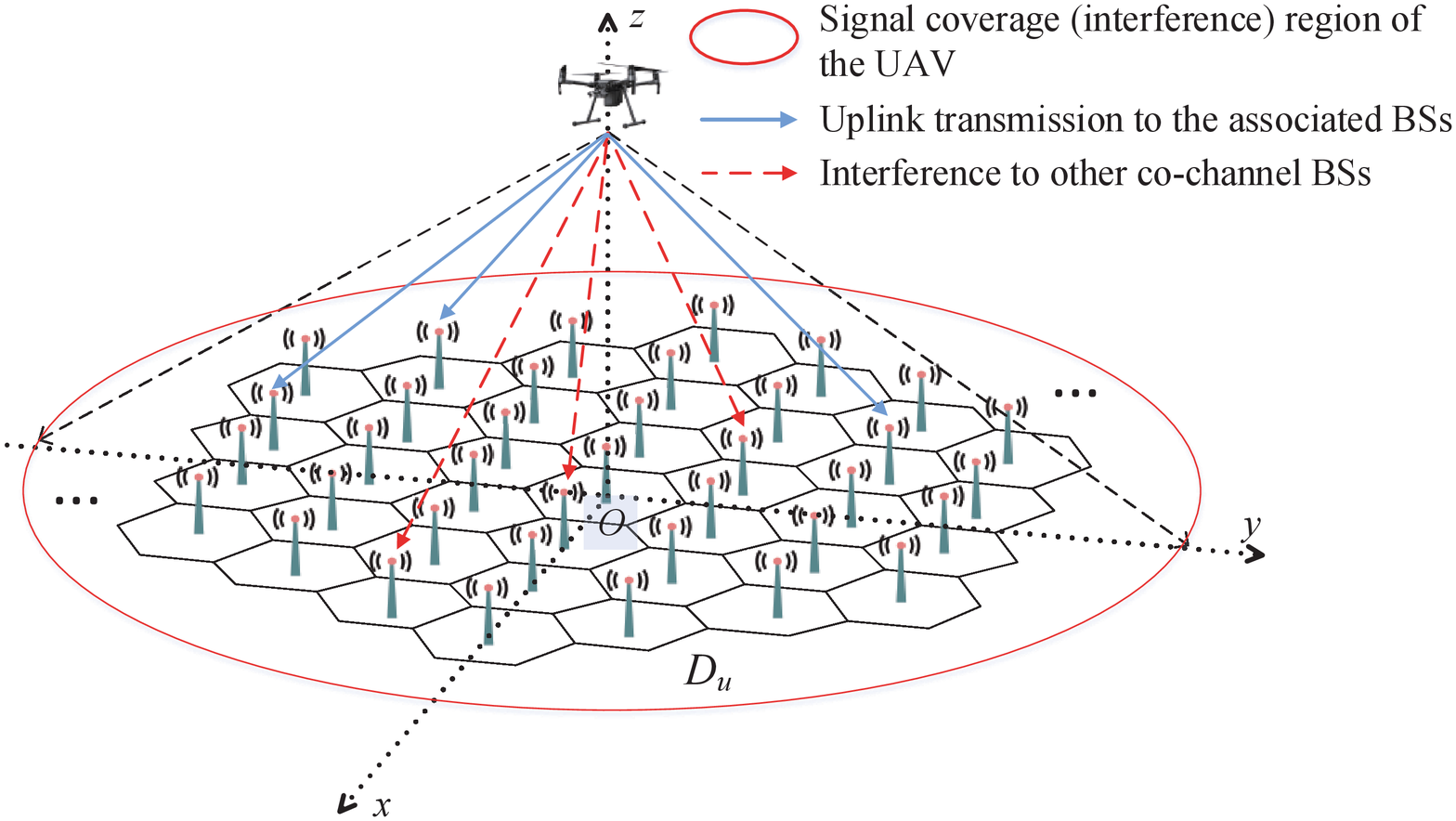}
\DeclareGraphicsExtensions.
\caption{Uplink UAV communication in a cellular network.}\label{up}
\vspace{-12pt}
\end{figure}
In this paper, we aim to investigate the uplink interference mitigation techniques for a cellular network with co-existing UAV and ground UEs, as shown in Fig.\,\ref{up}. Due to the strong A2G LoS channels, the UAV can be associated with multiple BSs in its signal coverage (as well as interference) region at the same time, but also generates severe interference to a large number of non-associated co-channel BSs. To mitigate the uplink interference from the UAV, one straightforward solution is to employ the orthogonal multiple access (OMA). With OMA, the UAV avoids causing any interference to the ground UEs in its coverage region by only transmitting in the resource blocks (RBs) that have not been assigned to any ground UEs in all the cells in the region. However, with increasing ground UE density and due to frequency reuse, the number of RBs available to the UAV decreases rapidly, and as a result the UAV's uplink transmission rate is severely limited. In contrast, non-orthogonal multiple access (NOMA) with interference cancellation (IC)\cite{saito2013non,ding2014performance,dai2015non,al2014uplink} is potentially a more appealing solution due to the following reasons. First, NOMA allows the UAV to reuse the RBs that have been assigned to ground UEs, which helps improve the UAV's rate performance at high ground UE density. Second, since the UAV at high altitude typically has stronger LoS channels with the BSs than ground UEs, each occupied BS (with served ground UEs in the uplink) can employ the IC to first decode the strong signal from the UAV and then subtract it for decoding the ground UE's signal\cite{ding2016impact}. However, for our considered system, conventional NOMA with local IC at individual BSs only, termed \emph{non-cooperative NOMA}, may provide very limited gain compared to the OMA. This is because in order to cancel the UAV's uplink interference, the UAV's signal at each RB needs to be decoded at all occupied BSs. As a result, the UAV's achievable rate is severely limited by the occupied BS with the worst channel condition with the UAV. Moreover, since the IC is only locally performed at each occupied BS, many unoccupied BSs with better channel condition are simply kept idle and not utilized to improve the system throughput. Owing to the above drawbacks of non-cooperative NOMA, innovative NOMA techniques are needed to mitigate the UAV's uplink interference more effectively yet without significantly compromising its achievable rate.

Motivated by the above, this paper proposes a new NOMA scheme, termed \emph{cooperative NOMA}, by exploiting the cooperative IC among the BSs via their backhaul links (e.g., the existing X2 link in LTE\cite{dahlman20134g}). Specifically, some BSs are first selected at each RB to decode the UAV's signal. Then the decoded UAV's signal is forwarded to their backhaul-connected BSs for IC. Since the selected BSs generally have better channel conditions than the occupied BSs, the UAV's rate performance can be greatly improved, as compared to the non-cooperative NOMA.
\vspace{-12pt}
\subsection{Main Contributions}
The main contributions of this paper are summarized as follows. First, for achieving a flexible performance trade-off between the UAV and ground UEs under our proposed cooperative NOMA, we aim to maximize their weighted sum-rate by jointly optimizing the UAV's uplink rate and power allocations over multiple RBs. We show that the UAV transmit rate and power at each RB have non-trivial effects to the cooperative IC and hence the achievable rates of the ground UEs. However, this optimization problem is non-convex and difficult to be solved optimally. Thus, we first study two special cases with the egoistic and altruistic transmission strategies of the UAV, respectively, under each of which the optimal solution can be efficiently obtained and important design insights are revealed.

Next, we consider the general case and show the main difficulty of solving our problem lies in that even with fixed rate allocations at RBs, the optimization of power allocations requires an exhaustive search of a large number of discrete power levels over all RBs, for which the complexity is prohibitively high in practice. As such, we propose an equivalent reformulation of the original problem, by replacing the rate allocations at different RBs with their BS associations. Then it is shown that the power allocation becomes a continuous optimization problem, which can be efficiently solved via the successive convex approximation (SCA) technique. By iteratively optimizing the power allocations and BS associations in an alternating manner, the overall algorithm is shown to converge both analytically and numerically. Last, based on the channel models recommended by the 3rd Generation Partnership Project (3GPP)\cite{3GPP36777,3GPP38901}, numerical results are provided to show the significant throughput gains of the proposed cooperative NOMA over the traditional OMA as well as the non-cooperative NOMA benchmark.
\vspace{-12pt}
\subsection{Related Work}
Despite that UAV communications in cellular networks have drawn increasing attention recently, there are only a handful of works \cite{zeng2018cellular,lin2018mobile,yaj2018interference,geraci2018supporting,amorim2018measured,zhang2018cellular,liu2018multi,cellular2018mei} that addressed the aerial-ground interference mitigation problem. In \cite{zeng2018cellular,lin2018mobile,yaj2018interference,geraci2018supporting,amorim2018measured}, the authors evaluated the performances of several existing techniques in LTE, such as three-dimensional (3D) beamforming, closed-loop power control, and massive multiple-input multiple-output (MIMO) for UAV communications via simulations or measurements. An interference-aware path planning design was proposed in \cite{zhang2018cellular} for a cellular-connected UAV under a given communication quality of service requirement with the ground BSs. The authors in \cite{liu2018multi} proposed a multi-beam UAV communication in cellular uplink, where the cooperative IC is applied jointly with transmit beamforming to mitigate the UAV's uplink interference to the ground UEs. In \cite{cellular2018mei}, inter-cell interference coordination (ICIC) designs were proposed for UAV uplink communication in the cellular network to maximize the network throughput by treating the interference as noise. However, none of the above works considers NOMA jointly with the cooperative IC for the UAV uplink communication, which motivates the current work.

On the other hand, NOMA has been extensively studied for terrestrial networks. Interested readers may refer to \cite{ding2017survey,islam2017power,huang2018signal,wan2018non,vaezi2018multiple} for the detailed literature survey. Most of the existing works on NOMA considered the single-RB and single-cell system setting, while some efforts have been made recently to address the more general multi-RB and/or multi-cell NOMA\cite{shin2017non,qian2017joint,zhang2017downlink,fu2017distributed,yang2018power,you2018resource}. However, the terrestrial NOMA techniques considered in the above works may be ineffective to mitigate the more severe interference as well as exploit the higher macro-diversity gain in the UAV's uplink communication due to the unique LoS-dominated channels with the ground BSs.

Finally, it is worth noting that NOMA has been applied to UAV-aided terrestrial communications \cite{sharma2017uav,sohail2018non,pan2018network,nguyen2018novel,nasir2018uav,rupasinghe2018non,hou2018multiple,wu2018uav}, where the UAV is deployed as an aerial BS to serve ground UEs via NOMA. In contrast, this paper focuses on investigating NOMA in the UAV uplink communication, where the UAV is considered as an aerial UE instead of the aerial BS in the above works.
\vspace{-10pt}
\subsection{Organization}
The rest of this paper is organized as follows. Section \Rmnum{2} presents the system model and Section \Rmnum{3} provides the problem formulation. Section \Rmnum{4} considers two special cases of the problem and solves them optimally to draw important insights. Section \Rmnum{5} considers the general case and proposes an efficient algorithm to obtain a suboptimal solution. Section \Rmnum{6} presents the simulation results on the performance of the proposed cooperative NOMA as compared to other benchmark schemes. Finally, Section \Rmnum{7} concludes the paper and points out directions for future work.\vspace{-6pt}

\section{System Model}
As shown in Fig.\,\ref{up}, we consider the uplink communication in a given subregion of the cellular network serving a UAV UE and a set of ground UEs. For simplicity, the shape of each cell is assumed to be hexagonal. For the purpose of exposition, we assume that the UAV is equipped with a single antenna, while each BS employs a fixed antenna pattern (see Section \ref{sim} for details). Due to the LoS-dominated A2G channel, the uplink signal from the UAV may interfere with the uplink transmissions from a large number of ground UEs using the same set of RBs at their associated BSs. Centered at the UAV's horizontal location on the ground, we consider there are in total $J$ BSs located in the UAV's signal coverage (or interference) region $D_u$, as shown in Fig.\,\ref{up}. For BSs outside this region, we assume that the signal strength from the UAV is attenuated to the level below the background noise and thus the resulted interference can be ignored. We assume that the total number of orthogonal RBs assigned for the UAV's uplink communication is $N$, which are denoted by the set ${\cal N}\triangleq\{1,2,\cdots,N\}$. For ease of reference, the main symbols used in this paper are listed in Table \ref{variable}.

In the sequel, we first describe the considered cellular network before and after the UAV UE is added. Then, we introduce the proposed cooperative NOMA scheme designed for mitigating the strong uplink interference from the UAV. In particular, the relationship between the ground UEs' sum-rate and the UAV's power and rate allocations is characterized under the proposed scheme.\vspace{-10pt}
\begin{table}[t]
\centering
\caption{List of Main Symbols}\label{variable}
\begin{tabular}{|c|l|}
\hline
Symbol & Description\\
\hline
$D_u$ & The UAV's signal coverage (or interference) region \\
\hline
${\cal J}$ & Set of all BSs in $D_u$ \\
\hline
$J$ & Number of BSs in $D_u$ \\
\hline
${\cal N}$ & Set of RBs assigned for the UAV's uplink communication \\
\hline
$N$ & Number of RBs assigned for the UAV's uplink communication \\
\hline
${\cal J}(n)$ & Set of all occupied BSs in RB $n$ \\
\hline
$K$ & Total number of UEs in $D_u$ \\
\hline
$k_j(n)$ & Index of the ground UE served by BS $j$ in RB $n$ \\
\hline
${\gamma}_{j}(n)$ & Receive SNR for ground UE $k_j(n)$\\
\hline
$\sigma_j^2(n)$ & \tabincell{l}{Total power of background noise and terrestrial ICI at cell $j$\\ in RB $n$} \\
\hline
$F_j(n)$ & \tabincell{l}{Channel power gain between the UAV and BS $j$ in RB $n$\\(normalized by $\sigma_j^2(n)$)} \\
\hline
$r_n$ & UAV's transmit rate in RB $n$\\
\hline
$p_n$ & UAV's transmit power in RB $n$\\
\hline
$R_j(n)$ & UAV's achievable rate in RB $n$ at BS $j$\\
\hline
$F_u(n)$ & Parameter, $F_u(n) \triangleq \mathop {\max}\nolimits_{j \in \cal J} \frac{F_j(n)}{1 + {\gamma_j}(n)}$\\
\hline
$M$ & Cancellation size \\
\hline
$N_j(M)$ & Set of the first $M$-tier neighboring BSs of BS $j$ \\
\hline
$C_j(M)$ & Parameter, $C_j(M)\triangleq \{j\} \cup N_j(M)$ \\
\hline
$\Lambda_n$ & Set of decodable BSs in RB $n$ \\
\hline
$\Omega_n$ & Set of cancelling BSs in RB $n$ \\
\hline
$j_n$ & Associated BS of the UAV in RB $n$ \\
\hline
\end{tabular}
\vspace{-6pt}
\end{table}

\subsection{Cellular Network with Ground UEs Only}
Assume that each BS $j \in {\cal J}\triangleq\{1,2,\cdots,J\}$ serves $K_j$ existing ground UEs over the $N$ RBs of our interest, with $K_j \ge 1, j \in \cal J$. Denote the total number of UEs in $D_u$ as $K=\sum\nolimits_{j=1}^J{K_j}$. Note that $N < K$ usually holds in practice due to the frequency reuse in the cellular network. In practice, inter-cell interference (ICI) exists if the ground UEs in different cells transmit in the same RB at the same time. In this paper, we assume that the ICI among ground UEs has been well mitigated by the existing ICIC techniques (see an example given in  Section \ref{sim}), such as cooperative RB allocation, beamforming, power control and so on. Thus, the terrestrial ICI is assumed to be much weaker and thus negligible as compared to the UAV's uplink interference at each BS.

For convenience, we define a set ${\cal J}(n) \subseteq \cal J$ for each RB $n \in {\cal N}$, where $j \in {\cal J}(n)$ if BS $j$ is currently serving a ground UE in RB $n$, and as a result ${\cal J}^c(n)={\cal J}\backslash {\cal J}(n)$. Let $k_j(n)$ be the index of the ground UE served by BS $j$ in RB $n$. Then we denote by $h_j(n)$ the fading channel between ground UE $k_j(n)$ and its serving BS (i.e., BS $j$) in RB $n$, which in general depends on the BS antenna gain, path-loss, shadowing, and small-scale fading. The ground UE $k_j(n)$'s transmit power is assumed to be $p_j(n)$. Then the received signal at BS $j$ in RB $n$ without the UAV's uplink transmission can be expressed as
\begin{equation}
y_{j,\text{w/o u}}(n)=h_j(n)x_j(n)+z_j(n),
\end{equation}
where $x_j(n)$ denotes the transmitted data symbol for ground UE $k_j(n)$ and satisfies ${\mathbb E}[{\lvert{x_j}(n)\rvert}^2] = p_j(n)$, and $z_j(n) \sim \mathcal{CN}(0,\sigma_j^2(n))$ comprises the background noise and terrestrial ICI at BS $j$ in RB $n$ (both assumed to be independently Gaussian distributed over $j$ and $n$ with $\sigma_j^2(n)$ denoting their total power). Accordingly, the receive signal-to-noise ratio (SNR) of ground UE $k_j(n)$ at its serving BS $j$ is given by
\begin{equation}
{\gamma}_{j}(n) = \frac{p_j(n)H_j(n)}{\sigma_j^2(n)},
\end{equation}
where $H_j(n)\triangleq \lvert h_j(n) \rvert^2$. Thus, the achievable sum-rate of all ground UEs in $D_u$ without the UAV's uplink transmission is given by
\begin{equation}\label{sum1}
R_{g,\text{w/o u}} = B\sum\limits_{n = 1}^{N}{\sum\limits_{j \in {\cal J}(n)}{\log_2 (1 + {\gamma}_{j}(n))}},
\end{equation}
in bits per second (bps), with $B$ denoting the total bandwidth per RB in Hertz (Hz). For notational convenience, we set $B=1$ Hz in the sequel of this paper, unless stated otherwise.
\vspace{-12pt}
\subsection{Cellular Network with New UAV UE Added}
Let $f_j(n)$ denote the channel between the UAV and BS $j$ in RB $n$. Due to the dominance of LoS propagation, we assume that the communication links from the UAV to BSs are frequency-flat over the spectrum of interest for simplicity. Thus, we have $f_j(n)=f_j, j \in {\cal J}, n \in {\cal N}$. For each RB $n \in \cal N$, suppose that the UAV's transmit power is $p_n$ with $p_n \ge 0$. Then the received signal at BS $j$ in RB $n$ becomes
\begin{equation}
y_{j,\text{w/ u}}(n)=h_j(n)x_j(n)+f_jx_u(n)+z_j(n),
\end{equation}
where $x_u(n)$ denotes the transmitted data symbol for the UAV in RB $n$, satisfying ${\mathbb E}[{\lvert{x_u}(n)\rvert}^2] = p_n$. The UAV's achievable rate in RB $n \in \cal N$ at BS $j \in \cal J$ can be expressed as
\begin{align}
R_j(n) &= {\log_2}\left(1 + \frac{p_n\lvert f_j \rvert^2}{\sigma_j^2(n) + p_j(n)H_j(n)}\right)\nonumber\\
&={\log _2}\left(1 + \frac{p_n{F_j}(n)}{1 + {\gamma_j}(n)}\right),\label{sum2}
\end{align}
where $F_j(n) \triangleq \lvert f_j \rvert^2/\sigma_j^2(n), j \in {\cal J}, n \in {\cal N}$. Notice that if $j \in {\cal J}^c(n)$, $\gamma_j(n)$ should be set to zero. For convenience, we refer to the parameter $\frac{F_j(n)}{1 + {\gamma_j}(n)}$ as the \emph{normalized} channel power gain between the UAV and BS $j$ in RB $n$. If the UAV's interference is treated as Gaussian noise at each \emph{occupied} BS (i.e., with served ground UEs), the achievable sum-rate of all ground UEs in RB $n$ after adding the UAV UE is expressed as
\begin{align}
R_{g,\text{w/ u}}(n) &=\sum\limits_{j \in {\cal J}(n)} {{\log_2}\left( {1 + \frac{{p_j}(n){H_j}(n)}{\sigma_j^2(n) + {p_n}\lvert f_j \rvert^2}} \right)}\nonumber\\
&= \sum\limits_{j \in {\cal J}(n)} {\log_2}\left(1 + \frac{{\gamma_j}(n)}{1 + p_n{F_j}(n)}\right).\label{sum3}
\end{align}
\vspace{-12pt}
\subsection{Cooperative NOMA}
To mitigate the UAV's uplink interference, in this paper we propose a new NOMA scheme with cooperative IC among BSs (termed ``cooperative NOMA''). Specifically, thanks to the superior UAV-BS LoS-dominated channels, at each RB there are in general a set of BSs (occupied or unoccupied) which are able to decode the UAV's uplink signal (even subjected to the interference from the co-channel ground UE in the case of occupied BSs), thus named as \emph{decodable BSs}. In practice, each decodable BS can differentiate the UAV's signal from ground UE's signal by applying the technique of \emph{interference detection}. Typical interference detection solutions can be categorized into two types: UE-based and network-based. Readers may refer to \cite{3GPP36777} for the detailed information. For each RB, one of the decodable BSs is appointed as the \emph{associated} BS of the UAV, which will forward the decoded UAV's message in that RB to the intended destination via the network routing. Moreover, by exploiting the backhaul links among BSs, the decodable BSs will also forward the decoded UAV's signals to their backhaul-connected BSs for IC. As a result, if an occupied BS is a decodable BS or backhaul-connected to one or more decodable BSs, it can completely cancel the interference from the UAV before decoding the ground UE's uplink signal. For convenience, we refer to the occupied BSs that are able to cancel the UAV's interference in each RB as the \emph{cancelling} BSs.

In this paper, we assume that each BS $j \in \cal J$ can forward the decoded UAV's signals to its first $M$ tiers ($M \ge 1$) of neighboring BSs. Let $N_j(M)$ be the set of the first $M$-tier neighboring BSs of BS $j \in \cal J$. Thus, if the UAV's signal is decoded by BS $j$, it will become available at all BSs in the set $C_j(M)\triangleq \{j\} \cup N_j(M)$ for IC. Notice that if $M=0$, the proposed scheme is reduced to the conventional NOMA with local IC at each BS only, thus termed ``non-cooperative NOMA'' in this paper. In contrast, with our proposed cooperative NOMA, when $M$ is sufficiently large, the decoded UAV's signals can be forwarded to any BSs in $D_u$, and thus the interference from the UAV can be cancelled at all occupied BSs in $D_u$, at the cost of higher complexity and possibly larger processing delay. For convenience, we denote $M$ as the ``cancellation size''.

Assume that the UAV transmits with rate $r_n$ in each RB $n \in {\cal N}$, with $r_n \ge 0$. If the UAV's signal in RB $n$ is decodable at BS $j$, i.e.,\vspace{-3pt}
\begin{equation}\label{Rjn}
R_j(n)={\log _2}\left(1 + \frac{p_n{F_j}(n)}{1 + {\gamma_j}(n)}\right) \ge r_n, n \in \cal N,
\end{equation}
then BS $j$ is a decodable BS in RB $n$. Let $\Lambda_n$ be the set of all decodable BSs in RB $n$, and $j_n \in \Lambda_n$ be the associated BS of the UAV in RB $n$. To maximally cancel the UAV's interference, each decodable BS $j \in \Lambda_n$ will decode (and forward) the UAV's signal in RB $n$ for IC if it is an occupied BS (or it has an $M$-tier neighboring BS that is occupied and needs the UAV's signal for IC), i.e., $C_j(M)\cap {\cal J}(n) \ne \emptyset$. According to (\ref{Rjn}), the following lemma holds.
\begin{lemma}\label{size}
Given fixed power allocations $\{p_n\}$, the size of the decodable BS set $\Lambda_n$ decreases with increasing $r_n$ for each $n \in \cal N$. On the other hand, given fixed rate allocations $\{r_n\}$, the size of $\Lambda_n$ increases with increasing $p_n$, for each $n \in \cal N$.
\end{lemma}

\begin{figure}[!t]
\centering
\includegraphics[width=2.6in]{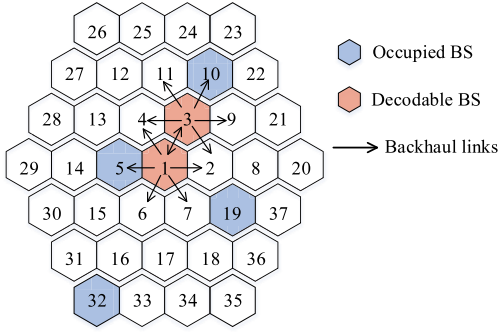}
\DeclareGraphicsExtensions.
\caption{An example of the proposed cooperative NOMA scheme for an RB with $J=37$ and $M=1$.}\label{example}
\vspace{-12pt}
\end{figure}

Lemma \ref{size} shows that the size of $\Lambda_n$ depends on both the UAV's power allocations $\{p_n\}$ and rate allocations $\{r_n\}$. To ensure that $\Lambda_n \neq \emptyset, \forall n \in \cal N$, it must hold that
\begin{equation}
r_n \le \mathop {\max}\limits_{j \in \cal J} R_j(n) = {\log _2}\left(1 + p_n F_u(n)\right), \forall n \in \cal N,
\end{equation}
where $F_u(n) \triangleq \mathop {\max}\limits_{j \in \cal J} \frac{F_j(n)}{1 + {\gamma_j}(n)}$. Let $\Omega_n$ denote the set of all cancelling BSs in each RB $n \in \cal N$. Notice that if the UAV's signal in RB $n$ is decoded by BS $j \in \Lambda_n$, then it will be cancelled by all occupied BSs in the set $C_j(M) \cap {\cal J}(n)$. Thus, $\Omega_n$ can be obtained as
\begin{equation}\label{cancel.set}
\Omega_n=\bigcup\limits_{j \in {\Lambda_n}} {\left(C_j(M)\cap {\cal J}(n)\right)},\; n \in \cal N.
\end{equation}
From (\ref{cancel.set}), it is easy to see that increasing the sizes of $\{\Lambda_n\}$ helps enlarge the sizes of $\{\Omega_n\}$ in general. For both non-cooperative and cooperative NOMA, the achievable sum-rate of all ground UEs in RB $n$ is expressed as
\begin{equation}\label{sum4}
R_g(n)\!=\!\sum\limits_{j \in \Omega_n}{\log_2\left(1\!+\!\gamma_j(n)\right)}+\!\sum\limits_{j \in \Omega^c_n}{\log_2}\left(1\!+\! \frac{{\gamma_j}(n)}{1 + {p_n}{F_j}(n)}\right),
\end{equation}
where $\Omega^c_n = {\cal J}(n) \backslash \Omega_n, n \in {\cal N}$. By comparing (\ref{sum4}) with (\ref{sum3}), it follows that fewer BSs will suffer from the UAV's interference in RB $n$ if $\Omega_n \ne \emptyset$, thanks to the local and cooperative IC at BSs. Notice that in the case of non-cooperative NOMA with $M=0$, the sets of cancelling BSs in (\ref{cancel.set}) are reduced to $\Omega_n={\Lambda_n} \cap {\cal J}(n), n \in \cal N$, which generally have smaller sizes than their counterparts in the case of cooperative NOMA with $M \ge 1$.
\begin{example}
An illustrative example of the proposed cooperative NOMA scheme for a given RB (say, RB $n'$ with $n' \in \cal N$) is shown in Fig.\,\ref{example}, where the total number of BSs is set to $J=37$, and the cancellation size is set to $M=1$. The occupied BSs in RB $n'$ include BSs 5, 10, 19 and 32, i.e., ${\cal J}(n')=\{5, 10, 19, 32\}$. Suppose that $r_{n'}$ is selected such that the UAV's signal in RB $n'$ is only decodable at BSs 1 and 3, i.e., $\Lambda_{n'}=\{1,3\}$. For BSs 1 and 3, we have $C_1(1)=\{1,2,3,4,5,6,7\}$ and $C_3(1)=\{1,2,3,4,9,10,11\}$, respectively. Then, according to (\ref{cancel.set}), the set of cancelling BSs in RB $n'$ is obtained as $\Omega_{n'}=\left(C_1(1) \cup C_3(1) \right) \cap {\cal J}(n')=\{5,10\}$, and thus $\Omega^c_{n'} = {\cal J}(n') \backslash \Omega_{n'}=\{19,32\}$. In contrast, if $M=0$, i.e., only non-cooperative NOMA is considered, then we have $\Omega_{n'}={\Lambda_{n'}} \cap {\cal J}(n')=\emptyset$, i.e., no occupied BS is able to cancel the UAV's interference in RB $n'$.
\end{example}

\section{Problem Formulation}
Under the proposed cooperative NOMA scheme, in order to achieve a flexible performance trade-off between the UAV and ground UEs in the region $D_u$, we aim to maximize their \emph{weighted} sum-rate, denoted by $Q(\{r_n\},\{p_n\})$, i.e.,
\begin{equation}\label{weight}
Q(\{r_n\},\{p_n\}) = \sum\limits_{n \in \cal N}{r_n} + \mu\sum\limits_{n \in \cal N}{R_g(n)},
\end{equation}
where $\mu \ge 0$ is a constant weight assigned to the ground UEs' sum-rate. To this end, we need to jointly design the rate allocations $\{r_n\}_{n \in \cal N}$ and the transmit power allocations $\{p_n\}_{n \in \cal N}$ for the UAV uplink communication. The design problem is formulated as
\begin{subequations}\label{op1}
\begin{align}
\nonumber \text{(P1)} \mathop {\max}\limits_{\{r_n\},\{p_n\}}&\; \sum\limits_{n \in \cal N}{r_n} + \mu\sum\limits_{n \in \cal N}{R_g(n)}\nonumber\\
\text{s.t.}\;\;&r_n \ge 0, \;p_n \ge 0, \forall n \in {\cal N},\label{op1a}\\
&r_n \le {\log _2}\left(1 + p_n F_u(n)\right), \forall n \in \cal N,\label{op1b}\\
&\sum\limits_{n \in \cal N}{p_n} \le P_{\max},\label{op1c}
\end{align}
\end{subequations}
where $P_{\max}$ denotes the maximum transmit power at the UAV. It is worth mentioning that there is in general a trade-off between maximizing the UAV's transmit rate and minimizing the UAV uplink interference (hence maximizing the ground UEs' sum-rate) in optimizing the rate allocations $\{r_n\}$ and power allocations $\{p_n\}$. On one hand, with any given power allocations, increasing transmit rates $\{r_n\}$ helps enhance the UAV's rate, but also shrinks the sizes of $\{\Lambda_n\}$ (and thus $\{\Omega_n\}$) according to Lemma \ref{size}. Consequently, the sum-rate of ground UEs, as given in (\ref{sum4}), may be degraded. On the other hand, with any given rate allocations, increasing $\{p_n\}$ helps enlarge the sizes of $\{\Lambda_n\}$ (and thus $\{\Omega_n\}$), which increases the first term in (\ref{sum4}). However, for those occupied BSs in $\{\Omega_n^c\}$, they will suffer from even stronger interference due to the increased $\{p_n\}$, which decreases the second term in (\ref{sum4}). As a result, the sum-rate of ground UEs may not change monotonically with $\{p_n\}$. To summarize, both $\{p_n\}$ and $\{r_n\}$ have non-trivial effects on the objective value of (P1).

It can be shown that the optimization problem (P1) is non-convex in general and thus difficult to be solved optimally. The main challenge lies in the non-trivial relationship between the weighted sum-rate and the UAV's power and rate allocations. To draw essential insights into our proposed cooperative NOMA scheme, we first study the optimal solution to problem (P1) in the two special cases with $\mu \rightarrow 0$ and $\mu \rightarrow +\infty$, respectively, in the next section.

\begin{remark}
Notice that we have assumed the transmit powers of ground UEs, i.e. $p_j(n)$'s in (P1), are given as fixed. This is because our main focus is on investigating how to efficiently integrate the UAV into a cellular network with existing terrestrial UEs. The proposed scheme is thus designed flexibly to be used with any terrestrial ICIC designs, which determines the terrestrial UEs' transmit powers in the uplink.
\end{remark}

\section{Special Cases}\label{special}
In this section, we focus on solving (P1) under two heuristic UAV transmission schemes, namely the egoistic scheme and the altruistic scheme, corresponding to the two special cases with $\mu \rightarrow 0$ and $\mu \rightarrow +\infty$ in (P1), respectively. Under both cases, we obtain the corresponding optimal solution to (P1), which helps reveal the effect of the cancellation size $M$ on the achievable rates of the proposed cooperative NOMA scheme.
\vspace{-20pt}
\subsection{Egoistic Scheme}
First, we consider the egoistic scheme by setting $\mu\rightarrow 0$ in (P1). When $\mu\rightarrow 0$, the UAV only aims to maximize its own achievable rate. Hence, the constraint (\ref{op1b}) must hold with equality, i.e., $r_n = {\log _2}\left(1 + p_n F_u(n)\right), \forall n \in \cal N$. This implies that, for each RB $n \in \cal N$, the UAV's signal is only decoded at the BS with the highest normalized channel power gain with the UAV among all BSs in $D_u$, which is thus the associated BS of the UAV in RB $n$ and denoted by
\begin{equation}\label{assoc1}
j_n^{\text{eg}}=\arg \mathop {\max }\limits_{j \in {\cal J}} {\frac{F_j(n)}{1 + {\gamma_j}(n)}}, n \in \cal N,
\end{equation}
which leads to $\Lambda_n=\{j_n^{\text{eg}}\}$ and $\Omega_n=C_{j_n^{\text{eg}}}(M) \cap {\cal J}(n)$. As a result, the optimal power allocations $\{p_n\}$ in this egoistic scheme should be water-filling over all RBs, denoted by
\begin{equation}\label{wf1}
p_n^{\text{eg}} = \left(\frac{1}{\lambda\ln 2}-\frac{1}{F_u(n)}\right)^+, n \in \cal N,
\end{equation}
where $\left(\cdot\right)^+\triangleq\max\{\cdot,0\}$ and $\lambda$ is a constant ensuring that $\sum\nolimits_{n \in {\cal N}}{p_n^{\text{eg}}} = P_{\max}$. The optimal UAV rate allocations $\{r_n\}$ are thus given by
\begin{equation}\label{wf2}
r_n^{\text{eg}} = {\log _2}\left(1 + p_n^{\text{eg}} F_u(n)\right), n \in \cal N.
\end{equation}

From (\ref{wf1}) and (\ref{wf2}), it is observed that the optimal UAV power and rate allocations in the egoistic scheme are regardless of $M$. Hence, the non-cooperative NOMA scheme with $M=0$ is able to achieve the same UAV rate performance as the cooperative NOMA. Nonetheless, the cooperative NOMA scheme in general yields a higher sum-rate of ground UEs, since the UAV's interference can be cancelled at more BSs with increasing $M$, which leads to the following proposition.
\begin{proposition}\label{ego}
The sum-rate of ground UEs under the egoistic scheme is monotonically non-decreasing with $M$, and eventually converges to the maximum sum-rate of ground UEs, i.e., $R_{g,\text{w/o u}}$ given in (\ref{sum1}), when $M \ge M_0$, where
\begin{equation}
M_0 = \arg \mathop {\min } M,\;\;\text{s.t.}\;\;{\cal J}(n) \subseteq C_{j_n^{\text{eg}}}(M), \forall n \in \cal N.
\end{equation}
\end{proposition}
\begin{IEEEproof}
In the egoistic scheme, the set of cancelling BSs in each RB $n$ is $\Omega_n=C_{j_n^{\text{eg}}}(M) \cap {\cal J}(n)$. Since increasing $M$ enlarges the size of $C_{j_n^{\text{eg}}}(M)$, the size of $\Omega_n$ must be monotonically non-decreasing with $M$. Then, from (\ref{sum4}), it follows that the sum-rate of ground UEs should also be monotonically non-decreasing with $M$. Eventually, if $M \ge M_0$, all occupied BSs in ${\cal J}(n)$ could receive the decoded UAV's signal in RB $n$ from BS $j_n^{\text{eg}}, n \in \cal N$. In this case, the maximum sum-rate of ground UEs, $R_{g,\text{w/o u}}$, can be achieved as if there was no interference from the UAV.
\end{IEEEproof}

Proposition \ref{ego} demonstrates that increasing the cancellation size $M$ helps improve the sum-rate of ground UEs in the egoistic scheme. In particular, when $M \ge M_0$, the UAV could attain its maximum transmit rate with its interference cancelled by all occupied BSs. Thus, in this case, the corresponding power and rate allocation solutions should be optimal for (P1) for any $\mu \ge 0$. However, if $M$ is small (e.g., $M=0$ for the non-cooperative NOMA case), each $\Omega_n$ may only contain very few or even no occupied BS. As a result, there may be a large number of occupied BSs outside $\{\Omega_n\}$ which are overlooked by the egoistic UAV transmission scheme and thus suffer from the UAV's uplink interference.
\vspace{-10pt}
\subsection{Altruistic Scheme}
Next, we consider the altruistic UAV transmission scheme. This corresponds to $\mu \rightarrow +\infty$ in (P1) under which the UAV needs to preserve the maximum sum-rate of ground UEs, i.e., $R_{g,\text{w/o u}}$ given in (\ref{sum1}). Apparently, if RB $n \in \cal N$ has not been occupied by any ground UEs in $D_u$, i.e., ${\cal J}(n)=\emptyset$, the UAV's signal in RB $n$ should still be decoded at BS $j_n^{\text{eg}}$ only, as given in (\ref{assoc1}). In contrast, if ${\cal J}(n) \ne \emptyset$, to preserve $R_{g,\text{w/o u}}$, we need to have $\Omega_n = {\cal J}(n)$. To this end, for each BS $j \in {\cal J}(n)$, a decodable BS should be available in the set $C_j(M)$. In order to maximize the UAV's transmit rate in RB $n$, this decodable BS should be selected as the one with the maximum normalized channel power gain with the UAV, denoted by
\begin{equation}\label{decBS}
\eta_j(n)=\arg \mathop {\max }\limits_{l \in C_j(M)} {\frac{F_l(n)}{1 + {\gamma_l}(n)}}, j \in {\cal J}(n).
\end{equation}
Then the set of decodable BSs in RB $n$ should be $\Lambda_n = \{\eta_j(n)\}_{j \in {\cal J}(n)}$, and the UAV's maximum transmit rate in RB $n$ is constrained by the decodable BS with the smallest normalized channel power gain with the UAV, which is given by
\begin{equation}\label{rn.alt}
r_n={\log_2}\left(1 + p_n T_u(n)\right),
\end{equation}
where
\begin{equation}\label{Tu}
T_u(n) \triangleq \mathop {\min }\limits_{j \in {\cal J}(n)} \mathop {\max }\limits_{l \in {C_j}(M)} \frac{F_l(n)}{1 + \gamma_l(n)}.
\end{equation}
\begin{example}
To illustrate the altruistic scheme more clearly, we use the same example as Fig.\,\ref{example} again. To preserve the maximum sum-rate of ground UEs in a given RB $n'$, the UAV's interference should be cancelled at all the four occupied BSs in the set ${\cal J}(n')=\{5, 10, 19, 32\}$. For BS 5, its decodable BS should be selected as the one with the maximum normalized channel gain with the UAV in the set $C_5(1)=\{1,4,5,6,13,14,15\}$, denoted by $\eta_5(n')$ as in (\ref{decBS}). Similarly, for BSs 10, 19 and 32, their decodable BSs in RB $n'$, i.e., $\eta_{10}(n')$, $\eta_{19}(n')$ and $\eta_{32}(n')$, should be selected from the sets $C_{10}(1)=\{3,9,10,11,22,23,24\}$, $C_{19}(1)=\{2,7,8,18,19,36,37\}$ and $C_{32}(1)=\{16,31,32,33\}$, respectively. Supposing that $\eta_5(n')=1$, $\eta_{10}(n')=3$, $\eta_{19}(n')=2$ and $\eta_{32}(n')=16$, then we have $\Lambda_{n'}=\{\eta_j(n')\}_{j \in {\cal J}(n')}=\{1,2,3,16\}$. Finally, the decodable BS with the smallest normalized channel gain with the UAV in the set $\Lambda_{n'}$ determines the UAV's maximum transmit rate in RB $n'$ as given in (\ref{rn.alt}) and (\ref{Tu}).
\end{example}

Accordingly, the optimal UAV power allocations in the above altruistic scheme, denoted by $p_n^{\text{al}}$, should also be water-filling over RBs, i.e.,
\begin{equation}\label{wf3}
p_n^{\text{al}}=
\begin{cases}
\left(\frac{1}{\lambda\ln 2}-\frac{1}{F_u(n)}\right)^+, & \text{if}\;{\cal J}(n)=\emptyset\\
\left(\frac{1}{\lambda\ln 2}-\frac{1}{T_u(n)}\right)^+, & \text{otherwise}
\end{cases}
\end{equation}
where $\lambda$ is a constant ensuring that $\sum\nolimits_{n \in {\cal N}}{p_n^{\text{al}}} = P_{\max}$.  The corresponding optimal rate allocations, denoted by $\{r_n^{\text{al}}\}$, are given by
\begin{equation}\label{wf4}
r_n^{\text{al}}=
\begin{cases}
{\log _2}\left(1 + p_n^{\text{al}} F_u(n)\right), & \text{if}\;{\cal J}(n)=\emptyset\\
{\log _2}\left(1 + p_n^{\text{al}} T_u(n)\right), & \text{otherwise}
\end{cases}
\end{equation}

Different from the egoistic scheme, the optimal power and rate allocations in the altruistic scheme are dependent on $M$. As a result, the non-cooperative NOMA with $M=0$ in general attains a smaller UAV's achievable rate than the cooperative NOMA with $M \ge 1$, since the decodable BSs in each RB can be selected from a larger set in the latter case with increasing $M$. Similar to Proposition \ref{ego}, we have the following proposition for the altruistic scheme.
\begin{proposition}\label{alt}
For the altruistic scheme, the UAV's transmit rate, i.e., $\sum\nolimits_{n \in \cal N} r_n^{\text{al}}$, is monotonically non-decreasing with $M$, and eventually converges to that by the egoistic scheme, i.e., $\sum\nolimits_{n \in \cal N} r_n^{\text{eg}}$ with $r_n^{\text{eg}}$'s given in (\ref{wf2}), when $M \ge M_0$.
\end{proposition}
\begin{IEEEproof}
For each RB $n \in \cal N$, increasing $M$ enlarges the size of $C_j(M)$ for each BS $j \in {\cal J}(n)$. Thus, $T_u(n)$ must be monotonically non-decreasing with $M$ according to (\ref{Tu}). As a result, $\sum\nolimits_{n \in \cal N} r_n^{\text{al}}$ with water-filling power allocations is also monotonically non-decreasing with $M$. When $M \ge M_0$, it follows that for each occupied BS $j \in {\cal J}(n)$, BS $j_n^{\text{eg}}$ is always available in the set $C_j(M)$, which leads to $T_u(n)=F_u(n)$ for each $n$ satisfying ${\cal J}(n) \ne \emptyset$. Thus, the UAV's maximum transmit rate, $\sum\nolimits_{n \in \cal N} r_n^{\text{eg}}$, can be achieved in this case.
\end{IEEEproof}

Proposition \ref{alt} demonstrates that increasing the cancellation size $M$ helps improve the UAV's transmit rate under the altruistic scheme. If $M \ge M_0$, the corresponding power and rate allocation solutions should also be optimal for (P1) for any $\mu \ge 0$. In this case, the altruistic scheme becomes equivalent to the egoistic scheme. However, if $M$ is small (e.g., $M=0$ for the non-cooperative NOMA case), it can be verified in this case that
\begin{equation}\label{oma}
T_u(n) = \mathop {\min }\limits_{j \in {\cal J}(n)} \frac{F_j(n)}{1 + \gamma_j(n)}.
\end{equation}
This reveals that the UAV's transmit rate in RB $n$ is severely limited by the occupied BS with the worst normalized channel power gain with the UAV in ${\cal J}(n)$. As such, the UAV's transmit rate under the altruistic scheme is practically low if non-cooperative NOMA is applied.

\section{Proposed Solution to (P1)}
In this section, we aim to solve problem (P1) in the general case with $0 < \mu < +\infty$ and $M < M_0$. To this end, we first propose an equivalent reformulation of (P1), and then solve the equivalent problem efficiently.
\vspace{-12pt}
\subsection{Equivalent Reformulation of (P1)}\label{reform}
As discussed in Section \Rmnum{2}, the main difficulty of solving (P1) lies in the coupled design of $\{r_n\}$ and $\{p_n\}$. To reduce complexity, a practical method is to alternatively optimize the rate allocations $\{r_n\}$ and the power allocations $\{p_n\}$ in an iterative manner until they both converge. However, even with fixed rate allocations, the power allocations optimization is still difficult to solve directly. The reason is that for each RB $n \in \cal N$, the size of $\Lambda_n$ (and $\Omega_n$) changes only when $p_n$ takes certain (up to $J$ in the worst case) discrete values. This renders the power allocation subproblem equivalent to a discrete optimization problem with $N$ discrete variables, each having up to $J$ discrete values. Moreover, the $N$ discrete variables are coupled in the sum power constraint (\ref{op1c}). As a result, an exhaustive search for all possible discrete power allocations at all RBs will incur a worst-case complexity in the order of ${\cal O}(J^N)$, which is prohibitive if $J$ and/or $N$ are practically large. To tackle this challenge, we reformulate the original problem (P1) into an equivalent form, in which the power allocation subproblem can be solved more efficiently. First, we have the following lemma.
\begin{lemma}\label{rn.lemma}
For (P1), with any given power allocations $\{p_n\}$, in each RB $n$, there must exist one decodable BS $j_n \in \cal J$ (regarded as the associated BS of the UAV in RB $n$), such that the optimal rate allocations $\{r_n^*\}$ satisfy
\begin{equation}\label{rn}
r_n^*=R_{j_n}(n)={\log _2}\left(1 + \frac{p_n{F_{j_n}}(n)}{1 + {\gamma_{j_n}}(n)}\right), \forall n \in \cal N.
\end{equation}
\end{lemma}
\begin{IEEEproof}
First, if ${\cal J}(n)=\emptyset$, i.e., RB $n$ has not been assigned to any ground UEs yet, the equality in (\ref{rn}) holds straightforwardly, and we have $j_n=j_n^{\text{eg}}$. On the other hand, if ${\cal J}(n) \ne \emptyset$, we can prove Lemma \ref{rn.lemma} by contradiction. Without loss of generality, suppose that the optimal transmit rate in an RB $n \in \cal N$, i.e., $r_n^*$, does not satisfy the equality in (\ref{rn}). It is easy to verify that
\begin{equation}\label{eq0}
r_n^* \ge \mathop {\min }\limits_{j \in \cal J} R_j(n).
\end{equation}
Since otherwise, we can always increase $r_n^*$ until the equality in (\ref{eq0}) holds, without decreasing the sum-rate of ground UEs. As a result, we can always find two BSs (named BS $a$ and BS $b$ with $a,b \in \cal J$), such that $R_a(n)<r_n^*<R_b(n)$ with
\begin{equation}\label{bound}
a = \arg \mathop {\max }\limits_{j \in {\cal J}, {R_j}(n) < r_n^*} R_j(n)
\end{equation}
and
\begin{equation}\label{bound}
b = \arg \mathop {\min }\limits_{j \in {\cal J}, {R_j}(n) > r_n^*} R_j(n).
\end{equation}
However, if we increase $r_n^*$ to $R_b(n)$, the sum-rate of ground UEs will not change, since the sets of decodable BSs are unchanged. This reveals that a larger objective value of (P1) can be achieved with $r_n=R_b(n)$, which contradicts our presumption. Lemma \ref{rn.lemma} is thus proved.
\end{IEEEproof}

According to Lemma \ref{rn.lemma}, it follows that the optimal UAV rate allocations $\{r_n^*\}$ can always be mapped into its unique BS associations $\{j_n\}$ via (\ref{rn}) with any given power allocations. Notice that the UAV's signal in each RB $n$ will be decoded at BS $j_n$, while it is also decodable at other BSs that have larger normalized channel power gain with the UAV than BS $j_n$. Based on the above, problem (P1) is equivalent to the following problem, in which BS associations $\{j_n\}_{n \in \cal N}$ and power allocations $\{p_n\}_{n \in \cal N}$ are jointly optimized, i.e.,
\begin{subequations}\label{op2}
\begin{align}
\nonumber \text{(P2)} \mathop {\max}\limits_{\{j_n\},\{p_n\}}&\; Q(\{j_n\},\{p_n\})\nonumber\\
\text{s.t.}\;\;&j_n \in {\cal J}, \forall n \in \cal N,\label{op2a}\\
&\sum\limits_{n \in \cal N}{p_n} \le P_{\max}, \;p_n \ge 0, \forall n \in \cal N,\label{op2b}
\end{align}
\end{subequations}
where
\begin{equation}
Q(\{j_n\},\{p_n\})\!=\!\sum\limits_{n \in \cal N}{\log_2\left(1 + \frac{p_n{F_{j_n}}(n)}{1 + \gamma_{j_n}(n)}\right)} + \mu\sum\limits_{n \in \cal N}{R_g(n)}.
\end{equation}
Moreover, in (P2), the sets of decodable BSs at different RBs are redefined as
\begin{equation}
\Lambda_n=\left\{j \in {\cal J}\left| {\frac{F_j(n)}{1 + {\gamma_j}(n)} \ge \frac{F_{j_n}(n)}{1 + {\gamma _{{j_n}}}(n)}}\right.\right\}, n \in \cal N.
\end{equation}
As will be shown later (see Section \ref{pw.alloc}), with the above problem reformulation, the power allocation subproblem becomes a continuous optimization problem and can be efficiently solved by the SCA algorithm.

Next, we will adopt an alternating optimization (AO) method to solve (P2), in which the power allocations $\{p_n\}_{n \in \cal N}$ and the BS associations $\{j_n\}_{n \in \cal N}$ are alternatively optimized in an iterative manner.
\vspace{-12pt}
\subsection{Power Allocation Optimization with Given BS Association}\label{pw.alloc}
First, we optimize the power allocations $\{p_n\}$ with fixed BS associations $\{j_n\}$ (and thus fixed $\{\Lambda_n\}$ and $\{\Omega_n\}$). In this case, problem (P2) is reduced to the following power allocation problem, i.e.,
\begin{subequations}\label{op3}
\begin{align}
\nonumber \mathop {\max}\limits_{\{p_n\}}&\; \sum\limits_{n \in \cal N}{{\log _2}\left(1 + \frac{p_n{F_{j_n}}(n)}{1 + {\gamma_{j_n}}(n)}\right)} + \mu\sum\limits_{n \in \cal N}R_g(n)\nonumber\\
\text{s.t.}\;\;&\sum\limits_{n \in \cal N}{p_n} \le P_{\max}, \label{op3a}\\
&p_n \ge 0, \forall n \in \cal N.\label{op3b}
\end{align}
\end{subequations}
Notice that problem (\ref{op3}) is a continuous optimization problem with respect to power allocations $\{p_n\}$, in contrast to its discrete counterpart for the original problem (P1) as discussed in Section \ref{reform}. However, problem (\ref{op3}) is still difficult to be optimally solved due to the non-concavity in its objective function, in which the second term of the ground UEs' sum-rate is not concave in $\{p_n\}$. To efficiently solve this problem and guarantee the convergence of the overall AO algorithm, we adopt the SCA technique to solve (\ref{op3}) locally optimally. The basic idea of the SCA is to approximate the non-concave objective function as a concave one given a local point in each iteration. By iteratively solving a sequence of approximated convex problems, we can obtain a locally optimal solution to (\ref{op3}).

Specifically, define $\{p_n^{(r)}\}$ as the given power allocation solution in the $r$-th SCA iteration. It can be shown that for each $n \in \cal N$, $R_g(n)$ is convex with respect to $p_n$. As such, the ground UEs' sum-rate $\sum\nolimits_{n \in \cal N}{R_{g}(n)}$ is a convex function of the UAV's power allocations $\{p_n\}$. By using the property that the first-order Taylor approximation of a convex function at any point is a global under-estimator of the convex function, we can obtain the following inequality (\ref{lb}), i.e.,
\begin{equation}\label{lb}
\sum\limits_{n \in \cal N}R_g(n) \ge A^{(r)}- \sum\limits_{n \in \cal N} {B_n^{(r)}(p_n - p_n^{(r)})},
\end{equation}
where
\begin{align}
A^{(r)} &=\sum\limits_{n \in \cal N}{\sum\limits_{j \in \Omega^c_n} {\log_2}\left(1 + \frac{\gamma_j(n)}{1 + p_n^{(r)}{F_j}(n)}\right)},\label{eq1}\\
B_n^{(r)} &= \sum\limits_{j \in \Omega^c_n}{\frac{F_j(n){\gamma_j}(n)}{\ln 2(1 + p_n^{(r)}{F_j}(n) + {\gamma_j}(n))(1 + p_n^{(r)}{F_j}(n))}}.\label{eq2}
\end{align}

With any given local point $\{p_n^{(r)}\}$ and the lower bound given in (\ref{lb}), problem (\ref{op3}) is approximated as the following problem in the $r$-th iteration of the SCA algorithm, i.e.,
\begin{align}
\mathop {\max}\limits_{\{p_n\}_{n \in {\cal N}}}&\; \sum\limits_{n \in \cal N}{\log_2\left(1 + \frac{p_n{F_{j_n}}(n)}{1 + {\gamma_{j_n}}(n)}\right)} - \mu\sum\limits_{n \in \cal N} {B_n^{(r)}p_n}\label{op4}\\
\text{s.t.}\;\;&\text{(\ref{op3a}),\;(\ref{op3b})},\nonumber
\end{align}
where some constant terms are omitted in the objective function of (\ref{op4}) for brevity.

Problem (\ref{op4}) is a convex optimization problem. By applying the Karush-Kuhn-Tucker (KKT) conditions (for which the details are omitted for brevity), the optimal solution to (\ref{op3}) can be obtained as
\begin{equation}\label{sol1}
p_n^{(r)*}=
\begin{cases}
{\tilde p}_n^{(r)}, &\text{if}\;\sum\nolimits_{n \in \cal N} {\tilde p_n^{(r)}\!\le\!P_{\max }}\\
{\left(\frac{1}{(\mu B_n^{(r)}\!+\!\nu)\ln 2}\!-\!\frac{1\!+\!{\gamma_{j_n}}(n)}{F_{j_n}(n)} \right)^+}, &\text{otherwise,}
\end{cases}
\end{equation}
for all $n \in {\cal N}$, where
\[{\tilde p}_n^{(r)} \triangleq {\left(\frac{1}{\mu B_n^{(r)}\ln 2} - \frac{1 + {\gamma_{j_n}}(n)}{{F_{j_n}}(n)}\right)^+},\]
and $\nu$ is a constant ensuring that $\sum\nolimits_{n \in {\cal N}}{p_n^{(r)*}} = P_{\max}$.

After solving problem (\ref{op3}) given any local point $\{p_n^{(r)}\}$, the SCA algorithm proceeds by iteratively updating $\{p_n\}$ based on the solution to problem (\ref{op4}). By applying the SCA convergence result in \cite{beck2010sequential}, it follows that a monotonic convergence is guaranteed for our proposed algorithm, since the objective value of problem (\ref{op3}) is non-decreasing over iterations. The proposed SCA algorithm to solve (\ref{op3}) is summarized in Algorithm \ref{Alg2}.
\begin{algorithm}
  \caption{SCA Algorithm for Solving Problem (\ref{op3})}\label{Alg2}
  \begin{algorithmic}[1]
    \State Initialize $\{p_n^{(1)}\}$. Let $r=1$.
    \State \textbf{Repeat}
    \State \quad Compute the optimal solution to problem (\ref{op4}) according to (\ref{sol1}) as $\{p_n^{(r)*}\}$.
    \State \quad Update $p_n^{(r+1)}=p_n^{(r)*}, n \in \cal N$.
    \State \quad Set $r=r+1$.
    \State \textbf{Until} the fractional increase of the objective value of problem (\ref{op3}) is below a threshold $\epsilon$.
  \end{algorithmic}
\end{algorithm}
\vspace{-18pt}
\subsection{BS Association Optimization with Given Power Allocation}
Given fixed power allocations $\{p_n\}$, note that problem (P2) can be decoupled into $N$ parallel BS association subproblems. The $n$-th subproblem, $n=1,2,\cdots,N$, is given by
\begin{align}
\nonumber \mathop {\max}\limits_{j_n}&\; {\log _2}\left(1 + \frac{p_n{F_{j_n}}(n)}{1 + {\gamma_{j_n}}(n)}\right) + \mu{R_g(n)}\nonumber\\
\text{s.t.}\;\;&j_n \in {\cal J}.\label{subprob}
\end{align}

Denote $j_n^*$ as the optimal solution to (\ref{subprob}). Depending on the cardinality of ${\cal J}(n)$, we consider the following two cases to obtain $j_n^*$, respectively.

\textbf{Case 1:} If ${\cal J}(n)=\emptyset$, it is easy to see that the optimal solution to problem (\ref{subprob}) should be the BS with the largest normalized channel power gain in $D_u$, i.e., $j_n^*=j_n^{\text{eg}}$.

\textbf{Case 2:} If $\lvert {\cal J}(n) \rvert \ge 1$, $j_n^*$ can be obtained via a full enumeration of the $J$ BSs. Nonetheless, a partial enumeration algorithm can be applied to obtain $j_n^*$ more efficiently, as summarized in Algorithm \ref{Alg1}.
\begin{algorithm}
  \caption{Partial Enumeration Algorithm for Solving Problem (\ref{subprob})}\label{Alg1}
  \begin{algorithmic}[1]
    \State Sort all BSs in $\cal J$ in the descending order based on their normalized channel power gains. Let ${\bm \pi}({\cal J})=\{b_1,b_2,\cdots,b_J\}$ denote the sequence of the sorted BSs with $b_i \in {\cal J}$, $1 \le i \le J$.
    \State Initialize $i=1$, $B=b_i$ and $\Omega_n=C_{b_i}(M) \cap {\cal J}(n)$. Compute the objective value of problem (\ref{subprob}) for $j_n=b_i$, denoted by $U(b_i)$.
    \State Set $i=i+1$.
    \While {$\Omega_n \subset {\cal J}(n)$}
    \If {$C_{b_i}(M) \cap {\cal J}(n) \ne \emptyset$}
    \State Update the set of cancelling BSs as ${\Omega_n} = {\Omega_n} \cup \left(C_{b_i}(M) \cap {\cal J}(n) \right)$ and compute $U(b_i)$.
    \State Update the best BS as $B\!=\!\arg \max \{U(b_i)\!,\!U(B)\}$.
    \EndIf
    \State Set $i=i+1$.
    \EndWhile
    \State Output the optimal BS association and the corresponding optimal value as $j_n^*=B$ and $U(B)$, respectively.
  \end{algorithmic}
\end{algorithm}

In Algorithm \ref{Alg1}, we define $B$ as the best associated BS of the UAV up to the current enumeration. Notice that $B$ is updated only if $C_{b_i}(M) \cap {\cal J}(n) \ne \emptyset$ (see line 7). This is because if $C_{b_i}(M) \cap {\cal J}(n)=\emptyset$, the set of cancelling BSs, as well as the sum-rate of ground UEs cannot be enlarged as compared to that for $j_n=b_{i-1}$. Since the UAV's achievable rate for $j_n=b_{i-1}$ is no smaller than that for $j_n=b_i$, we must have $U(b_i) \ge U(b_{i-1})$. Therefore, the optimal value of (\ref{subprob}) cannot be achieved at $j_n=b_i$. For a similar reason, the enumeration can terminate once the size of $\Omega_n$ reaches that of ${\cal J}(n)$ (see line 4). As such, the total search number in the partial enumeration algorithm can be significantly reduced as compared to its full version. Specially, if $M=0$, it is easy to verify that only the BSs in the set $\{b_1\} \cup {\cal J}(n)$ needs to be enumerated. Note that the above algorithm has the worst-case complexity of ${\cal O}(JN)$, which is polynomial.
\vspace{-8pt}
\subsection{Overall Algorithm and Convergence}
Based on the results presented in the previous two subsections, we propose an overall iterative algorithm for problem (P2) by applying the AO method. Specifically, the transmit power allocations $\{p_n\}$ and BS associations $\{j_n\}$ are alternately optimized, by solving problems (\ref{op3}) and (\ref{subprob}) respectively, while keeping the other one fixed. The details of the AO algorithm are summarized in Algorithm \ref{Alg3}. For simplicity, in this paper we set the initial BS associations $\{j_{n,1}\}$ and power allocations $\{p_{n,1}\}$ identical to those by the egoistic scheme, i.e., $j_{n,1}=j_n^{\text{eg}}$ and $p_{n,1}=p_n^{\text{eg}}$ for each $n \in \cal N$.
\begin{algorithm}
  \caption{AO Algorithm for Solving Problem (P2)}\label{Alg3}
  \begin{algorithmic}[1]
    \State Initialize $\{j_{n,1}\}$ and $\{p_{n,1}\}$. Let $m=1$.
    \State \textbf{Repeat}
    \State \quad Solve problem (\ref{op3}) with $j_n=j_{n,m}, n \in \cal N$ via Algorithm \ref{Alg2}, in which the power allocations are initialized as $p_n^{(0)}=p_{n,m}, n \in \cal N$. Denote the converged solution as $\{p_{n,m+1}\}$.
    \State \quad Solve problem (\ref{subprob}) with $p_n=p_{n,m+1}, n \in \cal N$ via Algorithm \ref{Alg1}, and denote the optimal solution as $\{j_{n,m+1}\}$.
    \State \quad Set $m=m+1$.
    \State \textbf{Until} the convergence condition is met.
    \State Output $\{j_{n,m}\}$ and $\{p_{n,m}\}$, and compute the corresponding rate allocation solution.
  \end{algorithmic}
\end{algorithm}

Next, we show that the proposed AO algorithm converges, as presented in the following proposition.
\begin{proposition}
Monotonic convergence is guaranteed for the proposed AO algorithm, i.e., $Q(\{j_{n,m+1}\},\{p_{n,m+1}\}) \ge Q(\{j_{n,m}\},\{p_{n,m}\})$.
\end{proposition}
\begin{IEEEproof}
First, in step 3 of Algorithm \ref{Alg3}, notice that the power allocations are initialized as $\{p_{n,m}\}$ in the SCA algorithm to solve problem (\ref{op3}). Since monotonic convergence is guaranteed in the SCA algorithm, it must hold that
\begin{equation}\label{eq3}
Q(\{j_{n,m}\},\{p_{n,m}\}) \le Q(\{j_{n,m}\},\{p_{n,m+1}\}).
\end{equation}
Second, in step 4 of Algorithm \ref{Alg3}, since the optimal solution of problem (\ref{subprob}) is obtained for given $\{p_{n,m+1}\}$, we have
\begin{equation}\label{eq4}
Q(\{j_{n,m}\},\{p_{n,m+1}\}) \le Q(\{j_{n,m+1}\},\{p_{n,m+1}\}).
\end{equation}
By combining (\ref{eq3}) and (\ref{eq4}), we obtain
\begin{equation}\label{eq5}
Q(\{j_{n,m}\},\{p_{n,m}\}) \le Q(\{j_{n,m+1}\},\{p_{n,m+1}\}),
\end{equation}
which demonstrates that the objective value of problem (P2) is non-decreasing after each iteration of Algorithm \ref{Alg3}. Since the objective value of problem (P2) is upper-bounded from above, Algorithm \ref{Alg3} is guaranteed to converge.
\end{IEEEproof}

\begin{remark}
In the traditional AO method, the subproblem for optimizing each block of variables is required to be \emph{optimally} solved in order to guarantee the convergence. Although in our case, the power allocation subproblem (\ref{op3}) is only \emph{locally} optimally solved, our analysis above has shown that the monotonic convergence is still guaranteed.
\end{remark}

\section{Numerical Results}\label{sim}
In this section, we provide numerical results to evaluate the performance of our proposed cooperative NOMA. An orthogonal frequency-division multiple access (OFDMA) system is considered. In order to mitigate the terrestrial ICI, the following terrestrial ICIC protocol is adopted: Each BS $j \in \cal J$ checks the availability of an RB in its first $q$ tiers ($q \ge 1$) of neighboring BSs (i.e., $N_j(q)$) before assigning it to a new ground UE. If an RB has been occupied by a ground UE in $N_j(q)$, BS $j$ cannot assign this RB to any new ground UE. By this means, the UEs associated with BS $j$ will not cause any uplink interference to all cells in $N_j(q)$.

Unless otherwise specified, the simulation settings are as follows. The tier of neighboring BSs is $q=1$ for the terrestrial ICIC\footnote{It can be verified via simulations that the terrestrial ICI attenuates to the level below background noise with high probability under $q=1$ and the considered settings.}. The total number of RBs in the subband that the UAV is allowed to access is $N=30$. Each RB consists of 12 consecutive OFDM subcarriers, with the subcarrier spacing being 15 kHz. The transmit powers of all active ground UEs are assumed to be identical as $23$ dBm. The cell radius is $800$ m, and the heights of BSs and UEs are set to be $H_B=25$ m and $H_{UE}=1.5$ m, respectively. The altitude of the UAV is fixed as 60 m. The carrier frequency $f_c$ is at $2$ GHz, and the noise power spectrum density at the receiver is $-164$ dBm/Hz including a 10 dB noise figure. For the terrestrial channels, the path-loss and shadowing are modeled based on the urban macro (UMa) scenario in the 3GPP technical report\cite{3GPP38901}. The small-scale fading is modeled as Rayleigh fading. The BS antenna pattern is assumed to be directional in the vertical plane but omnidirectional in the horizontal plane. Specifically, we consider in this paper a BS antenna pattern synthesized by a uniform linear array (ULA) with 10 co-polarized dipole antenna elements\cite{ballanis2016antenna}. The antenna elements are placed vertically with half-wavelength spacing and electrically steered with 10 degree downtilt angle. The ground UEs are all equipped with an isotropic antenna. On the other hand, the UAV-BS channels follow the probabilistic LoS/NLoS channel model based on the UMa scenario in the most recent 3GPP technical report\cite{3GPP36777}\footnote{Basically, there are three typical simulation scenarios for cellular-connected UAVs: rural macro (RMa), UMa, and urban micro (UMi)\cite{3GPP36777}. In the simulation, we select the urban environment with macro-cell to evaluate the performance. Notice that urban environment generally has larger non-LoS (NLoS) probability and path-loss exponent than the rural counterpart.}. We consider the network topology shown in Fig.\,\ref{example} with $J=37$. The BS in cell 1 is assumed to be located at the origin without loss of generality. The UAV's horizontal location is fixed at ${\bm q}_u=$(150 m, 420 m) in cell 1. The ground UEs' locations are randomly generated in the $J$ cells.

In the simulation, a non-orthogonal transmission scheme is included as a benchmark, where the UAV's signal can only be decoded at an unoccupied BS in each RB $n \in \cal N$. Mathematically, the associated design problem can be formulated by replacing $R_g(n)$ and $j_n \in {\cal J}$ in (P2) with $R_{g,\text{w/ u}}(n)$ and $j_n \in {\cal J}^c(n)$, respectively. Notice that the two special UAV transmission schemes, namely the egoistic scheme and the altruistic scheme, can also be considered in the above non-orthogonal transmission scheme. Particularly, in the altruistic non-orthogonal transmission scheme, the UAV can only access the available RBs that have not been assigned to any ground UEs in order to preserve $R_{g,\text{w/o u}}$. As a result, the altruistic non-orthogonal transmission scheme is essentially equivalent to the OMA scheme.

\begin{figure}[hbtp]
\centering
\includegraphics[width=3.4in]{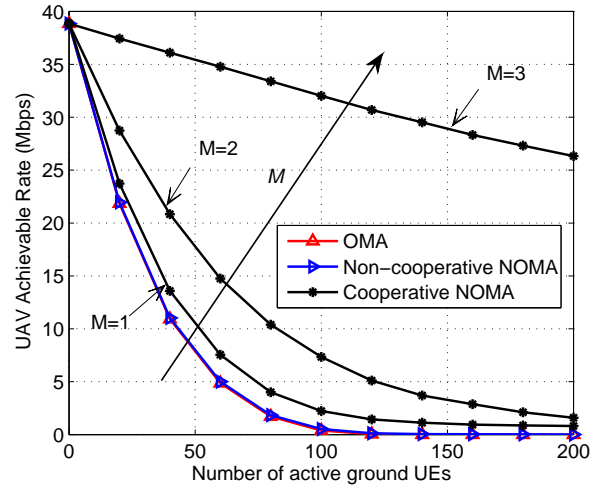}
\DeclareGraphicsExtensions.
\caption{UAV achievable rate versus number of active ground UEs in the altruistic scheme.}\label{Ds_thrpt_ic}
\vspace{-6pt}
\end{figure}
First, in Fig.\,\ref{Ds_thrpt_ic}, we plot the UAV's maximum achievable rate versus the number of active ground UEs $K$ in the altruistic scheme (i.e., when $\mu \rightarrow \infty$ or the maximum ground UEs' sum-rate $R_{g,\text{w/o u}}$ is preserved). The UAV's maximum transmit power $P_{\max}$ is set to 20 dBm. It is observed that the UAV's achievable rate degrades and gradually approaches zero with increasing $K$ with OMA. This is because as the number of active ground UEs increases, the total number of available RBs for OMA decreases and finally reaches zero, thus significantly degrading the UAV's rate performance at high ground UE density. Moreover, it is observed that the non-cooperative NOMA (with $M=0$) only provides marginal rate gain over the OMA. This is because for each RB $n$ satisfying ${\cal J}(n) \ne \emptyset$, the UAV's achievable rate is limited by the occupied BS with the worst normalized channel power gain with the UAV, as shown in (\ref{oma}). This renders the UAV's achievable rate in each RB $n$ practically very low, and thus results in only marginal rate gain over the OMA. In contrast, it is observed that the cooperative NOMA (with $M>0$) achieves increasingly more significant rate gains over the OMA as $M$ increases. This is because the cooperative NOMA scheme enables each occupied BS to receive the decoded UAV's signal from its nearby BSs that have higher normalized channel power gains. Such an observation is consistent with the result presented in Proposition \ref{alt}.

\begin{figure}[hbtp]
\centering
\includegraphics[width=3.4in]{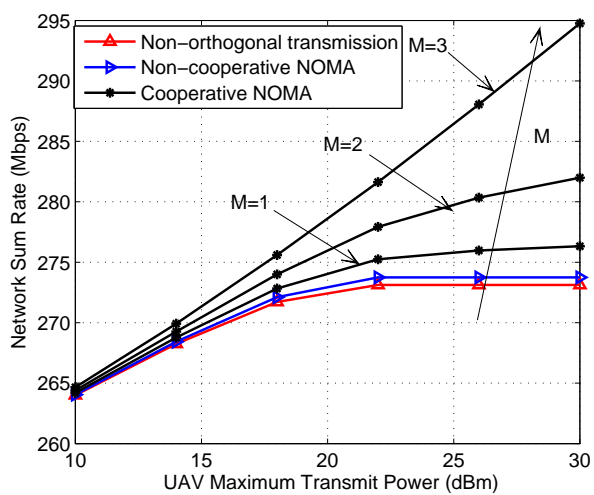}
\DeclareGraphicsExtensions.
\caption{Network sum-rate versus UAV maximum transmit power.}\label{Pw_thrpt_noma}
\vspace{-6pt}
\end{figure}
Next, by setting $\mu=1$, Fig.\,\ref{Pw_thrpt_noma} shows the network sum-rate after integrating the UAV into the network versus the UAV's maximum transmit power $P_{\max}$. The total number of active UEs is set to $K=150$. We first evaluate the performances of the two benchmark schemes, namely non-orthogonal transmission and non-cooperative NOMA with $M=0$. From Fig.\,\ref{Pw_thrpt_noma}, it is observed that the performance gap between non-cooperative NOMA and non-orthogonal transmission is not large. In addition, it is observed that the network sum-rates by the two benchmark schemes keep constant in the high UAV transmit power regime. This is because the rate loss of ground UEs increases with $P_{\max}$ due to the UAV's stronger uplink interference, and the UAV's achievable rate increase may not be sufficiently large to compensate for the rate loss of ground UEs in these two schemes. As a consequence, the UAV can only use a fraction of its maximum power budget in order to maximize the network sum-rate. In contrast, it is observed that the cooperative NOMA scheme offers significant sum-rate gains over the two benchmark schemes and the gains are more pronounced as $M$ or UAV transmit power increases, thanks to the reduced rate loss of ground UEs by more effective cooperative IC among the BSs.

\begin{figure}[hbtp]
\centering
\includegraphics[width=3.4in]{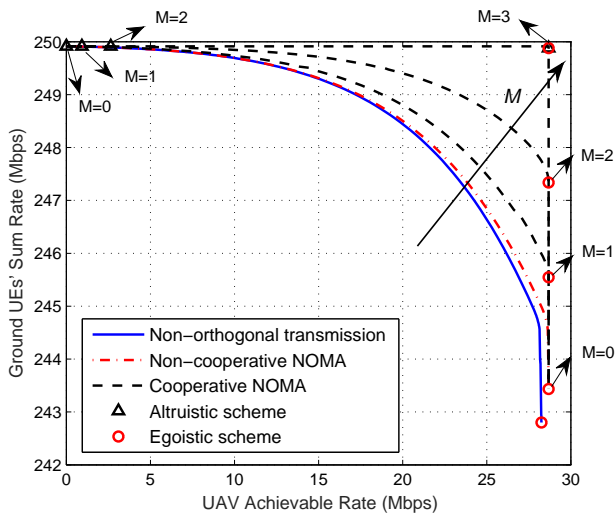}
\DeclareGraphicsExtensions.
\caption{Achievable rate region versus cancellation size $M$.}\label{pw.region}
\vspace{-6pt}
\end{figure}
Fig.\,\ref{pw.region} plots the achievable rate regions by different schemes with $P_{\max}=20$ dBm and $K=150$, which characterize the trade-off between the UAV's achievable rate and the ground UEs' sum-rate by varying the value of $\mu$. It is observed that both the UAV's achievable rate and the ground UEs' sum-rate are improved by applying IC as in non-cooperative and cooperative NOMA, as compared to those in the case of non-orthogonal transmission. This is because IC helps enhance the UAV's macro-diversity gain in BS association, and also reduces the rate loss of ground UEs due to the UAV's uplink interference. In addition, it is observed that increasing $M$ enhances the ground UEs' sum-rate and the UAV's achievable rate in the egoistic scheme and the altruistic scheme, respectively. Eventually, when $M=3$, these two schemes become equivalent and both achieve the maximum rates for the UAV and ground UEs (i.e., $r_n^{\text{eg}}$ and $R_{g,\text{w/o u}}$ as given in (\ref{wf2}) and (\ref{sum1}), respectively). This implies that with $M=3$, all occupied BSs can receive the decoded UAV's signals from the selected decodable BSs. As a result, the UAV's interference can be cancelled by all occupied BSs and thus there is no loss of ground UEs' sum-rate. The above results are in accordance with Propositions \ref{ego} and \ref{alt}. Finally, it is observed that as compared to the two benchmark schemes of non-orthogonal transmission and non-cooperative NOMA, the achievable rate regions by the proposed cooperative NOMA are dramatically enlarged with increasing $M$, especially when the UAV's achievable rate becomes large. This result indicates that the proposed cooperative NOMA is particularly beneficial when the rate demand of the UAV is high, which is usually the case for the considered uplink UAV communication in applications with high-rate payload data (e.g., high-resolution video).

\begin{figure}[hbtp]
\centering
\includegraphics[width=3.2in]{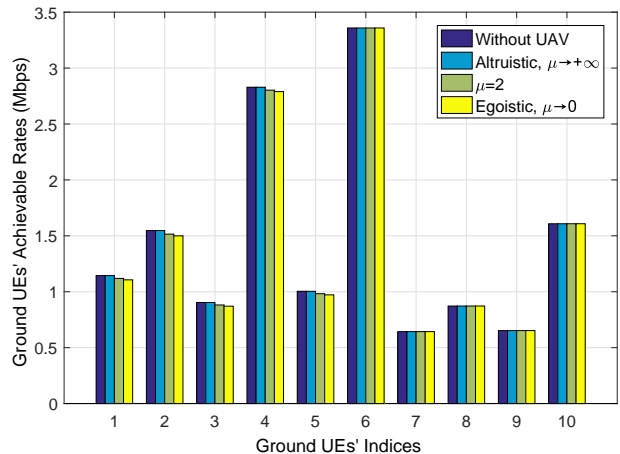}
\DeclareGraphicsExtensions.
\caption{Individual ground UEs' achievable rates.}\label{IndGrdRate}
\vspace{-6pt}
\end{figure}
Last, we show individual ground UEs' achievable rates under different weights in Fig.\,\ref{IndGrdRate}, with $P_{\max}=20$ dBm and $M=1$. Due to the space limit, we only choose 10 out of 150 ground users to show their achievable rates. From Fig.\,\ref{IndGrdRate}, it is observed that with the altruistic scheme, there is no rate loss for all ten ground UEs. However, with increasing $\mu$, the achievable rates of the first five ground UEs decrease. This is because the UAV imposes stronger uplink interference to the ground UEs, and the associated BSs of these five UEs cannot cancel the UAV interference with the optimized decodable BSs and $M=1$. On the contrary, for the last five ground UEs, it is observed that their achievable rates do not change with increasing $\mu$. This implies that their associated BSs are able to cancel the UAV interference, which validates the advantage of our proposed cooperative NOMA scheme.

\section{Conclusions}
This paper proposed a new cooperative NOMA scheme to mitigate the severe uplink interference due to the UAV's LoS channels with ground BSs in cellular-connected UAV communication. The proposed cooperative NOMA scheme includes the conventional non-cooperative NOMA with only local IC at individual BSs as a special case. Under the proposed scheme, we studied the weighted sum-rate maximization problem for the ground UEs and the UAV via jointly optimizing the UAV's uplink rate and transmit power allocations over multiple RBs. First, we obtained the optimal solutions to the formulated problem under the two special cases with egoistic and altruistic transmission strategies of the UAV, respectively, which reveal the effect of the cancellation size on the achievable rates. Next, we considered the general case and derived a locally optimal solution by introducing a judicious problem reformulation and applying the AO and SCA techniques.
Simulation results demonstrated that the proposed cooperative NOMA scheme yields higher achievable rates than the two benchmark schemes of non-orthogonal transmission and non-cooperative NOMA, especially when the ground traffic or the UAV's rate demand is high. It was also shown that increasing the cancellation size helps further improve the achievable rate trade-off between the UAV and ground UEs, at the cost of higher complexity and processing delay. This paper can be extended in several promising directions for future work, including UAV communication in the downlink, the more general case with multiple UAVs, as well as the practical design under imperfect channel knowledge and limited network coordination.

\bibliography{UAV_NOMA}
\bibliographystyle{IEEEtran}\vspace{-12pt}

\end{document}